\newcommand{\zlabel}[1]{\label{#1}} 
\newcommand{\smd}[1]{\sum_{#1}} 
\newcommand{\hH}{\hat{H}} 
\newcommand{\hJ}{\hat{J}} 
\newcommand{\hK}{\hat{K}} 
\newcommand{\ke}[1]{-\mbox{$\frac{1}{2}$}\nabla_{{#1}}^2} 
\newcommand{\roo}{{\rho_1}} 
\newcommand{\rc}[1]{r^{-1}_{#1}} 
\newcommand{\fc}{\frac} 
\newcommand{\lt}{\left} 
\newcommand{\rt}{\right} 
\newcommand{\ran}{\rangle} 
\newcommand{\lan}{\langle} 
\newcommand{\cE}{{\cal E}} 
\newcommand{\gm}{\gamma}
\newcommand{\si}{\sigma}
\newcommand{\dg}{\dagger} 
\newcommand{\mr}{\mathbf{r}}
\newcommand{\mz}{\mathbf{z}}
\newcommand{\mk}{\mathbf{k}}
\newcommand{\mx}{\mathbf{x}}
\newcommand{\om}{\omega}
\newcommand{\hE}{\hat{E}}
\newcommand{\dt}{\delta}
\newcommand{\al}{\alpha} 
\newcommand{\be}{\beta} 
\newcommand{\hs}[1]{\hspace{#1ex}}

\newcommand{\hvxc}{\hat{v}_{\text{xc}}} 
\newcommand{\hvcoroo}{\hat{v}_{\text{co}}^{\rho_1}} 
\newcommand{\vcoroo}{v_{\text{co}}^{\rho_1}} 
\newcommand{\tvcoroozvs}[1]{\tilde{v}_{\text{co}\roo}^{v_s\mathbf{z}_{#1}}} 
\newcommand{\eps}{\epsilon} 
\newcommand{\hF}{\hat{F}} 
\newcommand{\hv}{\hat{v}} 
\newcommand{\nn}{\nonumber} 
\newcommand{\rco}{r_{12}^{-1}} 
\newcommand{\Eco}{E_{\text{co}}} 
\newcommand{\cEco}{{\cal E}_{\text{co}}} 
\newcommand{\vep}{\varepsilon} 
\newcommand{\chF}{{\cal \hat{F}}} 
\newcommand{\ctF}{{\cal \tilde{F}}} 
\documentclass[preprint,prb,aps,amsmath]{revtex4}
\usepackage{graphicx}
\usepackage{dcolumn}
\usepackage{epic}%
\usepackage{eepic}%
\usepackage{bm}

\begin{document}


\title{The correlation energy as an explicit functional of the one-particle density
matrix from a determinantal reference state}

\author{James P. Finley}
\email{james.finley@enmu.edu}
\affiliation{
Department of Physical Sciences,
Eastern New~Mexico University,
Station \#33, Portales, NM 88130}
\affiliation{
Department of Applied Chemistry,
Graduate School of Engineering,
The University of Tokyo,
7-3-1 Hongo, Bunkyo-ku,
Tokyo, 113-8656 Japan}
\date{\today}


\begin{abstract}

Using an approach based on many body perturbation theory, the correlation energy
$\cEco$ is expressed as an explicit functional of $\rho_1$, $v$, and $v_s$, where
$\rho_1$ is the one-particle density matrix from the noninteracting, or reference,
determinantal-state; $v$ is the external potential from the interacting, or
target, state; $v_s$ is the (kernel of the) external potential from the
noninteracting determinantal-state. In other words we have $\cEco[\rho_1,v,v_s]$.
Anther possibility is the following explicit functional:
$\cEco[\rho_1,v_{\text{co}},v_s]$, where $v_{\text{co}}$ is the (kernel of the)
correlation potential from the noninteracting Hamiltonian.  The proposed method
can, in principle, be used to compute $\cEco$ in a very accurate and efficient
manner, since, like the Kohn--Sham approach, there are no virtual orbitals to
consider. However, in contrast to the Kohn--Sham approach, $\cEco$ is a known,
explicit functional that can be approximated in a systematic manner. For
simplicity, we only consider noninteracting closed-shell states and target states
that are nondegenerate, singlet ground-states; so, in that case, $\rho_1$ denotes
the spin-less one-particle density matrix from the determinantal reference state.

\end{abstract}

\maketitle

\section{Introduction}

The Kohn-Sham version of density functional theory has been very successful in the
description of electronic structure for quantum chemistry and condensed matter
physics.\cite{Dreizler:90,Parr:89,Springborg:97,Ellis:95,Gross:94,Seminario:95,Handy:97}
Unlike pure density functional approaches,\cite{Hohenberg:64a,Dreizler:90,Parr:89}
the Kohn--Sham method uses a set of occupied orbitals from a noninteracting state,
where this determinantal state shares its electron density $\rho$ with the target,
or interacting, state $|\Psi\ran$. In the Kohn--Sham approach, the kinetic energy
(through the) first order is not an explicit functional of $\rho$, but, instead,
this functional depends on the one-particle density matrix $\rho_1$ from the
noninteracting, determinantal state. The Kohn-Sham approach requires as input the
exchange-correlation functional $E_{\text{xc}}^{\text{KS}}$---or equivalently the
exchange $E_{\text{x}}$ and correlation-energy $\Eco^{\text{KS}}$
functionals---where $E_{\text{xc}}^{\text{KS}}$ is required to be an explicit
functional of $\rho$.  Unfortunately, $E_{\text{xc}}^{\text{KS}}$ is an unknown
functional, and there is no systematic method to improve approximations.

The optimized potential method
\cite{Fiolhais:03,Sharp:53,Talman:76,Li:93,Shaginyan,Goorling:94,Grabo:00} is a
density functional approach that can convert a nonlocal operator into a local one,
where the exchange-correlation functionals can depend on both the occupied and
virtual orbitals. Unfortunately, this method lacks the efficiency of other
Kohn--Sham approaches. Furthermore, the optimized potential method introduces
functional that---in contrast to many wave function methods---are not invariant to
a unitary transformation of either the occupied or virtual orbitals; the local
potentials also depend on the orbital energies.

High levels of approximations---beyond the Kohn--Sham approaches---can be obtained
by wave function methods,\cite{Lindgren:86,Paldus:75,Szabo:82,Harris:92} including
the coupled cluster method, many-body perturbation theory, and configuration
interaction. Often, however, these methods are much less efficient than the Kohn
Sham approaches, where, typically, wave function methods consider a large number
of 2-electron molecular integrals, depending on both the occupied and virtual
orbitals, and these integrals must be computed and utilized in calculations
involving large atomic-orbital basis sets.

In order to improve the efficiency of the wave function methods, especially in
regards to their scaling with molecular size, perturbative methods have been
developed based on localized molecular orbitals.\cite{Pulay:86,Saebo:87,Head-Gordon:98,Schultz:1999,Hetzer:2000,Finley:locmp2,Nakao:2004}
An alternative approach uses a Laplace transform to remove the energy denominators
in perturbation theory, yielding approaches involving correlation energy
expressions that depend explicitly on the atomic-orbital basis
set.\cite{Haser:1992,Scuseria:1999} This method has also been used with the
coupled cluster theory.\cite{Ayala:1999}

Our research interest is in the development of methods that bridge the gap between
wave function approaches and Kohn--Sham density functional theory. However, these
two different approaches already share some common features. For example, as
mentioned above, Kohn--Sham introduces a determinantal wave function and treats
the first-order kinetic energy in the same manner as in wave function methods,
i.e., as a functional of the one-body density matrix from a
determinant. Furthermore, hybrid density
functionals,\cite{Becke:93,Burke:97,Perdew:96,Ernzerhof:96} including
B3LYP,\cite{Becke:93,Stephens:94} introduce a component of the exact exchange---a
functional of $\roo$---even though these approaches violate the Hohenberg--Kohn
theorem\cite{Hohenberg:64a} by using the nonlocal exchange operator.

A simple generalization of the Kohn--Sham functionals involves using the exact
exchange energy ${\cal E}_x$ (with its corresponding non-local operator) and, in
addition, a correlation-energy functional that also depends on $\roo$. Such an
approach could probably be derived within a variant of the Kohn--Shan
scheme. However, it is also reasonable to base such a formalism on wave function
methods, since, for example, the correlation-energy, say $\cEco$, from wave
function methods is an implicit functional of $\roo$, and does not implicitly
depend on the orbitals, even though there is an explicit dependence.  (This is
easily proven by noting that we have $\cEco={\cal E} - E_1[\roo]$, where ${\cal
E}$ is the exact electron energy, and $E_1$ is the first order energy that is
determined by $\roo$.)  Furthermore, such a formalism can be based on any
reasonable orbitals: Hartree--Fock, Brueckner, and natural, but there is probably
no advantage to using Kohn--Sham orbitals. In addition, however, when considering
a wave function approach, $\cEco$ also has a natural dependence on the external
potential $v$ from the interacting or target state $|\Psi\ran$. (This is easily
proven from coupled cluster theory by noting the Fock operator $\hF$ appearance in
the coupled equations and noting that $\hF$ depends on $v$.) Therefore, it is
reasonable to have $\cEco$ depending on both on $v$ and $\roo$. Since $v$ is a
known function defining the molecular structure, the inclusion of the $v$ should
not produce any difficulties.

Below we introduce an approach based on many body perturbation
theory,\cite{Lindgren:86,Paldus:75,Raimes:72,Szabo:82} where the correlation
energy $\cEco$ is expressed as an explicit functional of $\rho_1$, $v$, and $v_s$,
where $\rho_1$ is the one-particle density matrix from the noninteracting, or
reference, state; $v$ is the external potential from the interacting state
$|\Psi\ran$; $v_s$ is the (kernel of the) external potential from the
noninteracting state. In other words we have $\cEco[\roo,v,v_s]$.  Here, the $v_s$
explicit dependence appears because the individual diagrams from the perturbative
expansion depend on the zeroth-order Hamiltonian, or, equivalently, on $v_s$,
since---together with the number of electrons---$v_s$ defines the zeroth-order
Hamiltonian. However, unless the perturbation expansion is truncated at some
order, it is easily proven that $\cEco$ does not implicitly depend on $v_s$, since
an exact $\cEco$ does not depend of the zeroth-order Hamiltonian. (Infinite order
summations, e.g., the coupled cluster method, also have $\cEco$ not depending on
$v_s$.)

This method presented below can, in principle, be used to compute $\cEco$ in an
accurate and efficient manner, since, like the Kohn--Sham approach, there are no
vitual orbitals to consider. However, in contrast to the Kohn--Sham approach,
$\cEco$ is a known, explicit functional that can be approximated in a systematic
manner. Furthermore, in contrast to the optimized potential method,
\cite{Fiolhais:03,Sharp:53,Talman:76,Li:93,Shaginyan,Goorling:94,Grabo:00} the
functionals are invariant to a unitary transformation of occupied orbitals.  The
proposed method, like the efficient wave function methods mentioned
above,\cite{Haser:1992,Scuseria:1999,Ayala:1999} removes the energy denominators
by using Laplace transforms, as well as other techniques, to obtain the
correlation energy as an explicit functional of $\rho_1$, $v$, and $v_s$.  For
simplicity, we only consider noninteracting closed-shell states and target states
that are nondegenerate, singlet ground-states; so, in that case, $\rho_1$ denotes
the spin-less one-particle density matrix from the determinantal reference state.

Let us also mention that we have recently derived a generalization of the
Kohn--Sham approach in which the correlation energy $\cEco$ is assumed to be an
explicit functional of $v$ and
$\roo$.\cite{Finley:ssks.arxiv,Finley:ssks,Finley:eecp.arxiv,Finley:eecp} In a
similar Brueckner-orbital method, called reference-state one-particle density
matrix theory,\cite{Finley:bdmt,Finley:bdmt.arxiv,Finley:B3lyp} $\cEco$ is also
assumed to be an explicit functional of $v$ and $\roo$, where for closed-shell
systems, the local density approximation (LDA) and the LYP, BLYP, and B3LYP
functionals was shown to be compatible with the method.\cite{Finley:B3lyp} In both
of these two approach, errors from Coulomb self-interactions do not occur, nor the
need to introduce functionals defined by a constraint search. By utilizing the
method presented here, explicit forms of these two correlation-energy functionals
can be obtained.\cite{Finley:tobe}

\section{The noninteracting state}

A determinantal state is a wave function that is an eigenfunction of a
noninteracting Hamiltonian, say $\hH_{Nv_s}$, where $\hH_{Nv_s}$ has no
electron-electron interaction terms.  These noninteracting Hamiltonian
$\hH_{Nv_s}$ are completely specified by the number of electrons $N$ and the
kernel of an external operator, say $v_s$. Explicitly we have
\begin{equation} \zlabel{5320}
\hH_{v_s} = \sum_{i=1}^N \chF_{v_s}(\mr_i),
\end{equation}
where 
\begin{equation} \zlabel{1827}
\chF_{v_s}  = \ke{} + \hv_s,
\end{equation}
and the nonlocal operator $\hv_s$---which we require to be spin-free---is defined
by its kernel $v_s$, i.e., we have
\begin{equation} 
\hv_s\chi(\mr_1) = \int d\mr_2 \, v_s(\mr_1,\mr_2) \chi(\mr_2).
\end{equation}

Using second quantization, our noninteracting Hamiltonian, given by
Eq.~(\ref{5320}), becomes
\begin{equation} \zlabel{9229}
\hH_{v_s} = \sum_{ij} (i|\chF_{v_s}|j)\hE_{ij},
\end{equation}
where the spin-adapted excitation operator is given by
\begin{equation} \zlabel{8732}
\hE_{ij}=\sum_{\si} a_{i\si}^\dg a_{j\si}, 
\end{equation}
and the one-electrons integrals are spin-free and are written in chemist's
notation;\cite{Szabo:82} these integrals use a spatial orbital set, say
$\{\chi\}$; this set has the following form:
\begin{equation}\zlabel{6752}
\psi_{i\si}(\mr,\om)= \chi_i(\mr)\si(\om); \;\; \si=\al,\be,
\end{equation}
where the spatial and spin coordinates, $\mr$ and $\om$, can be denoted
collectively by $\mx$, e.g.,
\begin{equation}
\psi_{i\si}(\mx) = \psi_{i\si}(\mr,\om).
\end{equation}

It is well known that there is a one-to-one mapping between determinantal states
and their one-particle density matrices,\cite{Blaizot:86,Parr:89} say $\gm$. For
a closed-shell state we have
\begin{equation} \zlabel{7234}
\gm(\mx_1,\mx_2) = 
\sum_{x\in\{\chi_o\}}
\sum_\si \chi_{x}(\mr_1)\si(\om_1) \, \chi_{x}^*(\mr_2) \si^*(\om_2)
= 
\sum_{x\in\{\chi_o\}}
\chi_{x}(\mr_1) \chi_{x}^*(\mr_2)\dt_{\om_1\om_2},
\end{equation}
and the spin-less one-particle density-matrix is
\begin{equation} \zlabel{7298}
\roo(\mr_1,\mr_2) = 
\sum_{\om}  \gm(\mr_1,\om,\,\mr_2,\om) = 
2 \sum_{x\in\{\chi_o\}}  \chi_{x}(\mr_1) \chi_{x}^*(\mr_2),
\end{equation}
where the sum is over the occupied orbitals from Eq.~(\ref{6752}); this set of
orbitals is denoted by $\{\chi_o\}$. Comparing Eqs.~(\ref{7234}) and (\ref{7298})
yields
\begin{equation}
\gm(\mx_1,\mx_2) = \fc12 \roo(\mr_1,\mr_2) \dt_{\om_1\om_2}.
\end{equation}
Since our closed-shell determinantal states are determined by $\roo$, henceforth,
these kets are denoted by $|\roo\ran$.

Using Eq.~(\ref{9229}) and the determinantal state $|\roo\ran$, our eigenvalue
problem can be written as
\begin{equation} \zlabel{7539}
\hH_{v_s}|\roo\ran= 
2\lt( 
\mbox{\small $\displaystyle \sum_{w\in\{\chi_o\}}$}\eps_w\rt) |\roo\ran
= \cE_0|\roo\ran,
\end{equation}
where the set of occupied orbitals $\{\chi_o\}$ is defined by the lowest $N$
eigenstates of $\chF_{v_s}$,
\begin{equation} \zlabel{2190}
\chF_{v_s}\chi_w = \eps_w \chi_w, \;\; \chi_w\in \{\chi_o\} \longrightarrow \roo,
\end{equation}
and the right side indicates that these orbitals, from the set $\{\chi_o\}$,
determine $\roo$. The unoccupied, or virtual, orbital set $\{\chi_u\}$ are also
obtained from $\chF_{v_s}$:
\begin{equation} \zlabel{2194}
\chF_{v_s}\chi_r = \eps_r \chi_r, \;\; \chi_r\in \{\chi_u\} \longrightarrow \roo,
\end{equation}
where these orbitals also determine $\roo$, since, for a complete set of orbitals,
the unoccupied set determines the occupied set; two sets of orbitals that differ by
a unitary transformation are considered equivalent.

While there is no restriction placed upon $v_s$, normally it contains the external
potential $v$ from the interacting state $|\Psi_{Nv}\ran$, the Coulomb potential
$\hJ$, and an additional term, the exchange-correlation potential $\hvxc$, that we
also permit to depend on the one-particle density matrix:
\begin{equation} \zlabel{0097}
\hv_s = v + \hJ_{\rho_s}  + \hvxc^{\roo} = v + \hJ_{\rho_s} + \hK_{\roo} + \hvcoroo,
\end{equation}
and the latter expression defines the correlation potential $\hvcoroo$, where
$\hK$ is the exchange operator, defined below; the Coulomb potential $\hJ$, given
by
\begin{equation} \zlabel{4211}
\hJ_{\rho_s}(\mr_1)
= \int r_{12}^{-1} \rho_s(\mathbf{r}_2)d\mr_2,
\end{equation}
depends on the electron density $\rho_s$ of the
noninteracting state $|\roo\ran$:
\begin{equation}\zlabel{5267}
\rho_s(\mr) = \roo(\mr,\mr)
= 2\hs{-1} 
\sum_{w\in\{\chi_o\}}  \chi_{w}(\mr) \chi_{w}^*(\mr), 
\;\; \roo \longrightarrow \rho_s,
\end{equation}
and we use Eq.~(\ref{7298}); the right side of this equation indicates that
$\rho_s$ is determined by $\roo$.

The non-local operator $\hvxc$ can be identified as the exchange-correlation
operator. If $|\roo\ran$ is the Kohn-Sham noninteracting state, this operator is
local and it implicitly depends only on $\rho_s$, the density from $|\roo\ran$ and
the interacting state $|\Psi_{Nv}\ran$. For Brueckner orbital
theory,\cite{Finley:bdmt,Finley:bdft} this operator is nonlocal and may depend on
the external potential $v$. For Hartree--Fock theory, the correlation portion of
$\hvxc$ is absent; so, in that case, $\hvxc^{\roo}$ is just the exchange operator
$\hK$, where the kernel of this operator is
\begin{equation} \zlabel{2391}
K_{\roo}(\mr_1,\mr_2) = - \frac12 r_{12}^{-1} \rho_1(\mathbf{r}_1,\mathbf{r}_2).
\end{equation}
For latter use, we also write down the expression for the Fock operator:
\begin{equation} \zlabel{9158}
\hF_{\roo} = \ke{}  + v + \hJ_{\rho_s} + \hK_{\roo}. 
\end{equation}

\section{The interacting state}

Now consider an interacting system of electrons.  The electronic energy, say
$\cE_{Nv}$, for an eigenstate of an interacting Hamiltonian is determined by the
number of electrons $N$ and the one-body external potential $v$. This must be the
case, since the non-relativistic, electronic Hamiltonian is entirely specified by
$N$ and $v$:
\begin{equation} \zlabel{5278}
\hH_{Nv}= \sum_i^N \ke{i} + \sum_i^N v(i) + \fc12 \sum_{i\ne j}^N \sum_j^N r_{ij}^{-1}, 
\end{equation}
where $r_{ij}^{-1}$ is the electron-electron interaction operator. Furthermore,
since each $N$ and $v$ yields a different ground state wave function
$|\Psi_{Nv}\ran$.\cite{Dreizler:90} Therefore, the ground state energies
$\cE_{Nv}$, given by
\begin{equation}
\cE_{Nv} = \fc{\lan\Psi_{Nv} |H_{Nv}|\Psi_{Nv}\ran}{\lan\Psi_{Nv}|\Psi_{Nv}\ran},
\end{equation}
are unique functionals of $N$ and $v$.

Using a second quantization approach, the Hamiltonian does not depend on $N$, and
it can be expressed by
\begin{equation} \zlabel{0115}
\hH_v = 
\smd{ij}(i|(\ke{})|j) \hE_{ij} + \smd{ij}(i|v|j) \hE_{ij} + 
\fc12 \smd{ijkl} (ij|kl) \hE_{ijkl},
\end{equation}
where the symmetry-adapted excitation operators are given by Eq.~(\ref{6752}) and
\begin{equation}
\zlabel{7282}
\hE_{ijkl}=
\sum_{\si\lambda}
a_{i\si}^\dg a_{k\lambda}^\dg a_{l\lambda} a_{j\si},
\end{equation}
and the two electrons integrals are spin-free and are written in chemist's
notation.\cite{Szabo:82}

\section{The correlation energy as in explicit functional of 
\mbox{\large \lowercase{$v$}}, \mbox{\large $\roo$} and \mbox{\large \lowercase{$v_s$}}}

Using many body perturbation
theory,\cite{Lindgren:86,Paldus:75,Raimes:72,Szabo:82} we partition our
Hamiltonian into the zeroth order part and a perturbation $V$:
\begin{equation} \zlabel{0735}
\hH_v = \hH_{v_s} + V_\roo,
\end{equation}
where we choose $\hH_{v_s}$, from Eq.~(\ref{9229}), as our zeroth-order Hamiltonian.
Using Eqs.~(\ref{1827}), (\ref{9229}),  (\ref{0115}), and (\ref{0735}), we have
\begin{equation} \zlabel{9271}
V_\roo 
= 
 \fc12 \smd{ijkl} (ij|kl) \hE_{ijkl}
+ \smd{ij}(i|v|j) \hE_{ij}
- \smd{ij}(i|v_s|j) \hE_{ij}.
\end{equation}

The energy through the first order---the zeroth-order energy plus the correction
from the 1st order ---for many body perturbation theory is independent of the
zeroth-order Hamiltonian. This energy functional, given by
\begin{equation} 
E_1[\roo,v] = \lan \roo|H_{v}|\roo \ran,
\end{equation}
depends on the two functions: the external potential $v$ and the one-particle
density matrix $\roo$. Explicitly, this functional is given by
\begin{eqnarray} \label{4217}
E_1[\roo,v] = \int d\mr_1\, \lt[\ke{1}\roo(\mr_1,\mr_2)\rt]_{\mr_2=\mr_1} 
\hs{48}\\
\nn \hs{0.5}  \mbox{}
+   \int d\mr_1\, v(\mr_1)\rho(\mr_1) 
+   \fc12 \int \int \rco d\mr_1 d\mr_2 \rho(\mr_1)\, \rho(\mr_2) 
-   \fc14 \int \int \rco d\mr_1 d\mr_2 \roo(\mr_1,\mr_2)\, \roo(\mr_2,\mr_1).
\end{eqnarray}

In many-body perturbation theory, the electronic energy is given by an
order-by-order expansion
\begin{equation}
\cE = \cE_0 + \cE_1 + \cE_2 + \cE_3 + \cdots,
\end{equation}
where the energy through the first order is 
\begin{equation}
E_1 = \cE_0 + \cE_1,
\end{equation}
and the correlation energy $\cEco$ includes everything beyond the first order:
\begin{equation} \zlabel{0191}
\cEco = \cE - E_1 = \cE_2 + \cE_3 + \cdots.
\end{equation}

In order to obtain the an order-by-order expansion of the electronic energy $\cE$,
it is convenient to write the perturbation $V_\roo$, given by Eq.~(\ref{9271}), in
normal ordered form\cite{Lindgren:86,Paldus:75} using $|\roo\ran$ as the vacuum
state:
\begin{equation} 
V_\roo =  V_c^\roo + V_1^\roo + V_2^\roo,
\end{equation}
where the constant, one-, and two-body terms are given by
\begin{eqnarray} 
V_c^\roo &=&  \cE_1 =  E_1 - \cE_0, \\
\zlabel{7682}
V_1^\roo &=&  
-\smd{ij} (i|\hvcoroo|j) \{\hE_{ij}\}_{\roo}, 
\\
\zlabel{0285}
V_2^\roo &=& 
 \fc12 \smd{ijkl} (ij|kl) \{\hE_{ijkl}\}_{\roo},
\end{eqnarray}
where the correlation potential is defined by Eq.~(\ref{0097}), i.e.,
\begin{equation} \zlabel{5279}
\hvcoroo = v_s - v- \hJ_{\rho_s} - \hK_{\roo}, 
\end{equation}
and $\cE_0$, $E_1$, $\hJ$, and $\hK$ are given by Eqs.~(\ref{7539}), (\ref{4217}),
(\ref{4211}), and (\ref{2391}), respectively; the brackets $\{\cdots\}_\roo$
indicate that the excitation operators are in normal ordered form with respect to
the vacuum state $|\roo\ran$. Furthermore, subtracting Eq.~(\ref{9158}) from
(\ref{1827}) and using Eq.~(\ref{5279}), we have
\begin{equation} \zlabel{6483}
\hvcoroo =  \chF_\roo - \hF_\roo.
\end{equation}

Eq.~(\ref{7682}) is the most convenient expression for $V_1$. However, we can also
substitute Eq.~(\ref{5279}) into Eq.~(\ref{7682}), yielding individual diagrams
from the perturbation expansion that depend explicitly on $v$ and
$v_s$. Substituting Eq.~(\ref{6483}) into Eq.~(\ref{7682}) separates $V_1$ into
the zeroth-order part $\chF$ and the Fock-operator part $\hF$. (Note that
$\{\hF\}_{\roo}$ is the one-body portion of the
Hamiltonian.)\cite{Finley:bdmt.arxiv} While the individual terms from the
perturbation expansion can depend on $v_s$ or $\chF$, the correlation energy
$\cEco$, unless approximated from a truncated expansion, cannot depend on either
$v_s$ or $\chF$; we now prove that $\cEco$ is an implicit functional of $v$ and
$\roo$.

Consider the electronic energy $\cE$, determined by $v$ and $N$, while $E_1$ is an
explicit functional of $v$ and $\roo$.  Hence, the correlation energy, given by
($\cEco = \cE - E_1$), is determined by $v$, $\roo$, and $N$, or just $v$ and
$\roo$, since $\roo$ determines $N$. Therefore, as mentioned above, at least for
the ground state, the correlation energy is an implicit functional of $v$ and
$\roo$, i.e, $\cEco[v,\roo]$.  We now demonstrate that the individual perturbation
terms, or diagrams, can be written as explicit functional of $v$, $v_s$ and
$\roo$.

Generally speaking, perturbation theory expresses $\cEco$ as a functional of the
occupied and unoccupied orbitals from $|\roo\rangle$, their orbitals energies
$\{\eps\}$, and the one-body portion or the perturbation $V_1$, which, in our case
is $-\hvcoroo$, ($v - v_s + \hJ_{\rho_s} + \hK_{\roo}$), or ($\hF-\chF)$), as
indicated in Eqs.~(\ref{7682}), (\ref{5279}), and (\ref{6483}). The
correlation-energy diagrams, however, do not depend on the constant portion of the
perturbation $V_c$; also, the two-body part $V_2$ is universal, since this operator is
the same for all electronic systems.

Consider for the second-order (correction to the) energy, given by
\begin{equation} \zlabel{8581}
\cE_2 = \cE_{2u} + \cE_{2d} + \cE_{2x}, 
\end{equation}
where the Goldstone diagrams for these terms are given by the first three entries
of Fig.~\ref{9180}, respectively. For the first term, we have
\begin{equation}
\cE_{2u}[\chi_o,\chi_u,\eps,\vcoroo] = 2\smd{w\in\{\chi_o\}}\smd{r\in\{\chi_u\}}
\fc{1}{\eps_w-\eps_r} (r|\hvcoroo|w) (w|\hvcoroo|r),
\end{equation}
where this diagram depends explicitly on $\vcoroo$---the kernel of
$\hvcoroo$---the orbital energies \{$\eps\}$, and both the occupied and unoccupied
orbitals, $\{\chi_o\}$ and $\{\chi_u\}$. Writing out the integrations explicitly,
we have
\begin{equation}
\cE_{2u}[\chi_o,\chi_u,\eps,\vcoroo] = 
2\smd{wr}\int \int d\mr_1 d\mr_2 \, 
\fc{1}{\eps_w-\eps_r}
\hvcoroo(\mr_1)\chi_w(\mr_1)\chi_w^*(\mr_2) \hvcoroo(\mr_2)\chi_r(\mr_2)\chi_r^*(\mr_1).
\end{equation}
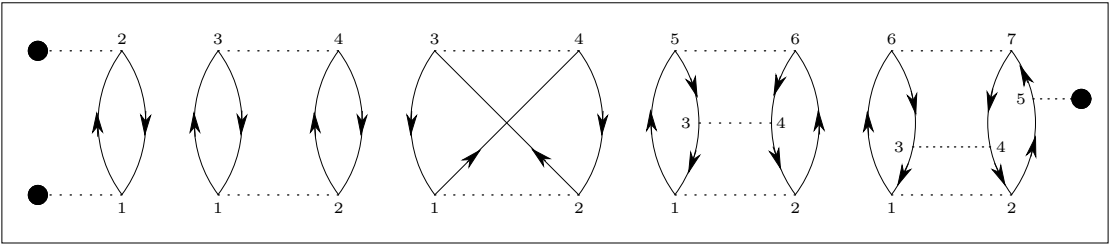
\begin{figure*}
\setlength{\unitlength}{0.00083333in}
\begingroup\makeatletter\ifx\SetFigFont\undefined%
\gdef\SetFigFont#1#2#3#4#5{%
  \reset@font\fontsize{#1}{#2pt}%
  \fontfamily{#3}\fontseries{#4}\fontshape{#5}%
  \selectfont}%
\fi\endgroup%
{\renewcommand{\dashlinestretch}{30}
\begin{picture}(6924,1539)(0,-10)
\put(237,312){\blacken\ellipse{120}{120}}
\put(237,312){\ellipse{120}{120}}
\put(237,1212){\blacken\ellipse{120}{120}}
\put(237,1212){\ellipse{120}{120}}
\put(6747,912){\blacken\ellipse{120}{120}}
\put(6747,912){\ellipse{120}{120}}
\path(12,1512)(6912,1512)(6912,12)
	(12,12)(12,1512)
\path(5665,507)(5602,377)
\blacken\path(5627.336,498.071)(5602.000,377.000)(5681.329,471.904)(5638.633,452.591)(5627.336,498.071)
\path(6205,510)(6269,377)
\blacken\path(6189.934,472.124)(6269.000,377.000)(6244.000,498.140)(6232.577,452.692)(6189.934,472.124)
\path(6413,1013)(6365,1117)
\blacken\path(6442.526,1020.617)(6365.000,1117.000)(6388.048,995.473)(6400.201,1040.731)(6442.526,1020.617)
\path(5686,960)(5713,820)
\blacken\path(5660.819,932.148)(5713.000,820.000)(5719.733,943.510)(5697.093,902.480)(5660.819,932.148)
\path(6193,974)(6166,834)
\blacken\path(6159.267,957.510)(6166.000,834.000)(6218.181,946.148)(6181.907,916.480)(6159.267,957.510)
\path(6433,554)(6459,674)
\blacken\path(6462.909,550.369)(6459.000,674.000)(6404.270,563.074)(6441.213,591.905)(6462.909,550.369)
\path(4870,1053)(4830,939)
\blacken\path(4841.423,1062.165)(4830.000,939.000)(4898.039,1042.299)(4857.811,1018.262)(4841.423,1062.165)
\path(4343,589)(4297,455)
\blacken\path(4307.588,578.239)(4297.000,455.000)(4364.337,558.758)(4324.274,534.449)(4307.588,578.239)
\path(4823,613)(4869,479)
\blacken\path(4801.663,582.758)(4869.000,479.000)(4858.412,602.239)(4841.726,558.449)(4801.663,582.758)
\path(4309,1042)(4349,928)
\blacken\path(4280.961,1031.299)(4349.000,928.000)(4337.577,1051.165)(4321.189,1007.262)(4280.961,1031.299)
\path(5419,684)(5412,830)
\blacken\path(5447.712,711.574)(5412.000,830.000)(5387.781,708.701)(5416.023,746.096)(5447.712,711.574)
\path(5112,691)(5105,837)
\blacken\path(5140.712,718.574)(5105.000,837.000)(5080.781,715.701)(5109.023,753.096)(5140.712,718.574)
\path(4065,691)(4058,837)
\blacken\path(4093.712,718.574)(4058.000,837.000)(4033.781,715.701)(4062.023,753.096)(4093.712,718.574)
\path(3762,836)(3755,690)
\blacken\path(3730.781,811.299)(3755.000,690.000)(3790.712,808.426)(3759.023,773.904)(3730.781,811.299)
\path(2570,836)(2563,690)
\blacken\path(2538.781,811.299)(2563.000,690.000)(2598.712,808.426)(2567.023,773.904)(2538.781,811.299)
\path(3402,524)(3322,597)
\blacken\path(3430.864,538.275)(3322.000,597.000)(3390.421,493.953)(3384.050,540.380)(3430.864,538.275)
\path(2915,517)(2995,590)
\blacken\path(2926.579,486.953)(2995.000,590.000)(2886.136,531.275)(2932.950,533.380)(2926.579,486.953)
\path(2261,836)(2254,690)
\blacken\path(2229.781,811.299)(2254.000,690.000)(2289.712,808.426)(2258.023,773.904)(2229.781,811.299)
\path(1969,678)(1962,824)
\blacken\path(1997.712,705.574)(1962.000,824.000)(1937.781,702.701)(1966.023,740.096)(1997.712,705.574)
\path(1511,837)(1504,691)
\blacken\path(1479.781,812.299)(1504.000,691.000)(1539.712,809.426)(1508.023,774.904)(1479.781,812.299)
\path(1218,678)(1211,824)
\blacken\path(1246.712,705.574)(1211.000,824.000)(1186.781,702.701)(1215.023,740.096)(1246.712,705.574)
\path(611,678)(604,824)
\blacken\path(639.712,705.574)(604.000,824.000)(579.781,702.701)(608.023,740.096)(639.712,705.574)
\path(911,836)(904,690)
\blacken\path(879.781,811.299)(904.000,690.000)(939.712,808.426)(908.023,773.904)(879.781,811.299)
\put(162.000,762.000){\arc{1500.000}{5.6397}{6.9267}}
\put(1362.000,762.000){\arc{1500.000}{2.4981}{3.7851}}
\put(4962.000,762.000){\arc{1500.000}{5.6397}{6.9267}}
\put(5712.000,762.000){\arc{1500.000}{5.6397}{6.9267}}
\put(6912.000,762.000){\arc{1500.000}{2.4981}{3.7851}}
\put(3612.000,762.000){\arc{1500.000}{5.6397}{6.9267}}
\put(4812.000,762.000){\arc{1500.000}{2.4981}{3.7851}}
\put(5562.000,762.000){\arc{1500.000}{2.4981}{3.7851}}
\put(4362.000,762.000){\arc{1500.000}{5.6397}{6.9267}}
\put(6162.000,762.000){\arc{1500.000}{2.4981}{3.7851}}
\put(3312.000,762.000){\arc{1500.000}{2.4981}{3.7851}}
\put(3012.000,762.000){\arc{1500.000}{5.6397}{6.9267}}
\put(762.000,762.000){\arc{1500.000}{5.6397}{6.9267}}
\put(1512.000,762.000){\arc{1500.000}{5.6397}{6.9267}}
\put(1962.000,762.000){\arc{1500.000}{2.4981}{3.7851}}
\put(2712.000,762.000){\arc{1500.000}{2.4981}{3.7851}}
\path(2712,312)(3614,1214)
\path(2712,1212)(3614,310)
\dottedline{60}(312,1212)(762,1212)
\dottedline{60}(312,312)(762,312)
\dottedline{60}(5562,1212)(6312,1212)
\dottedline{60}(5562,312)(6312,312)
\dottedline{45}(5694,612)(6180,612)
\dottedline{45}(6687,912)(6444,912)
\dottedline{60}(4212,1212)(4962,1212)
\dottedline{60}(4212,312)(4962,312)
\dottedline{60}(4362,762)(4812,762)
\dottedline{60}(2712,1212)(3612,1212)
\dottedline{60}(2712,312)(3612,312)
\dottedline{60}(1362,1212)(2112,1212)
\dottedline{60}(1362,312)(2112,312)
\put(732,1257){\makebox(0,0)[lb]{{\SetFigFont{6}{7.2}{\rmdefault}{\mddefault}{\updefault}2}}}
\put(732,197){\makebox(0,0)[lb]{{\SetFigFont{6}{7.2}{\rmdefault}{\mddefault}{\updefault}1}}}
\put(5532,197){\makebox(0,0)[lb]{{\SetFigFont{6}{7.2}{\rmdefault}{\mddefault}{\updefault}1}}}
\put(6282,197){\makebox(0,0)[lb]{{\SetFigFont{6}{7.2}{\rmdefault}{\mddefault}{\updefault}2}}}
\put(5532,1257){\makebox(0,0)[lb]{{\SetFigFont{6}{7.2}{\rmdefault}{\mddefault}{\updefault}6}}}
\put(6282,1257){\makebox(0,0)[lb]{{\SetFigFont{6}{7.2}{\rmdefault}{\mddefault}{\updefault}7}}}
\put(4182,197){\makebox(0,0)[lb]{{\SetFigFont{6}{7.2}{\rmdefault}{\mddefault}{\updefault}1}}}
\put(4932,197){\makebox(0,0)[lb]{{\SetFigFont{6}{7.2}{\rmdefault}{\mddefault}{\updefault}2}}}
\put(4182,1257){\makebox(0,0)[lb]{{\SetFigFont{6}{7.2}{\rmdefault}{\mddefault}{\updefault}5}}}
\put(4932,1257){\makebox(0,0)[lb]{{\SetFigFont{6}{7.2}{\rmdefault}{\mddefault}{\updefault}6}}}
\put(4249,727){\makebox(0,0)[lb]{{\SetFigFont{6}{7.2}{\rmdefault}{\mddefault}{\updefault}3}}}
\put(4843,727){\makebox(0,0)[lb]{{\SetFigFont{6}{7.2}{\rmdefault}{\mddefault}{\updefault}4}}}
\put(5579,577){\makebox(0,0)[lb]{{\SetFigFont{6}{7.2}{\rmdefault}{\mddefault}{\updefault}3}}}
\put(6339,877){\makebox(0,0)[lb]{{\SetFigFont{6}{7.2}{\rmdefault}{\mddefault}{\updefault}5}}}
\put(6213,577){\makebox(0,0)[lb]{{\SetFigFont{6}{7.2}{\rmdefault}{\mddefault}{\updefault}4}}}
\put(2682,1257){\makebox(0,0)[lb]{{\SetFigFont{6}{7.2}{\rmdefault}{\mddefault}{\updefault}3}}}
\put(2682,197){\makebox(0,0)[lb]{{\SetFigFont{6}{7.2}{\rmdefault}{\mddefault}{\updefault}1}}}
\put(3582,1257){\makebox(0,0)[lb]{{\SetFigFont{6}{7.2}{\rmdefault}{\mddefault}{\updefault}4}}}
\put(3582,197){\makebox(0,0)[lb]{{\SetFigFont{6}{7.2}{\rmdefault}{\mddefault}{\updefault}2}}}
\put(1332,197){\makebox(0,0)[lb]{{\SetFigFont{6}{7.2}{\rmdefault}{\mddefault}{\updefault}1}}}
\put(2082,197){\makebox(0,0)[lb]{{\SetFigFont{6}{7.2}{\rmdefault}{\mddefault}{\updefault}2}}}
\put(1332,1257){\makebox(0,0)[lb]{{\SetFigFont{6}{7.2}{\rmdefault}{\mddefault}{\updefault}3}}}
\put(2082,1257){\makebox(0,0)[lb]{{\SetFigFont{6}{7.2}{\rmdefault}{\mddefault}{\updefault}4}}}
\end{picture}
}
\caption{\zlabel{9180} The three diagrams from the second-order correlation-energy
${\mathcal E}_{\mathrm{co}}$ and two other diagrams.}
\end{figure*}

In order to remove the energy denominator, we use the following Laplace
transform:\cite{Haser:1992,Harris:92}
\begin{equation} \zlabel{6281}
\fc{1}{\eps_w-\eps_r} = 
\int_0^{\infty} dt  \, e^{-t(\eps_w-\eps_r)},
\end{equation}
giving
\begin{eqnarray}
\cE_{2u}[\chi_o,\chi_u,\eps,\hvcoroo] = 
\hs{65}\\\nn \hs{10} 
2\smd{wr} 
\int \int d\mr_1 d\mr_2 
\int_0^{\infty} dt_1   \, 
e^{t_1\eps_r} \chi_r^*(\mr_1) \hvcoroo(\mr_1) e^{-t_1\eps_w}\chi_w(\mr_1)
\chi_w^*(\mr_2) \hvcoroo(\mr_2)\chi_r(\mr_2).
\end{eqnarray}
Using Eqs.~(\ref{2190}) and (\ref{2194}), and the observation that $\chF_{\hv_s}$
depends explicitly on the kernel $v_s$, and, according to Eq.~(\ref{5279}),
$\hvcoroo$ depends explicitly on $v$ and $v_s$, we have
\begin{eqnarray}
\cE_{2u}[\chi_o,\chi_u,v,v_s] = 
\hs{69}\\\nn \hs{4} 
2\smd{wr} 
\int \int d\mr_1 d\mr_2 
\int_0^{\infty} dt_1   \, 
\lt(e^{t_1\chF_{v_s}^{1}} \chi_r^*(\mr_1)\rt) 
\hvcoroo(\mr_1)
\lt(e^{-t_1\chF_{v_s}^{1}}\chi_w(\mr_1)\rt)
\chi_w^*(\mr_2) \hvcoroo(\mr_2)\chi_r(\mr_2),
\end{eqnarray}
where $\chF_{v_s}^{i}$ denotes $\chF_{v_s}(\mr_i)$ and the round brackets
$\lt(\cdots \rt)$ serve to remind us that $e^{t_1\chF_{v_s}^{1}}$ acts exclusively
upon $\chi_r^*(\mr_1)$ and $e^{-t_1\chF_{v_s}^{1}}$ acts upon $\chi_w(\mr_1)$.
Note that, since $\hv_s$ and $\ke{}$ do not commute, the exponential operator,
defined by
\begin{equation} \label{7293}
e^{t\chF_{v_s}}=
\exp\lt[t(\ke{} + \hv_s)\rt] 
= 1 + t\lt(\ke{} + \hv_s\rt) + \fc{t^2}{2!}\lt(\ke{} + \hv_s\rt)\lt(\ke{} + \hv_s\rt) + \cdots, 
\;\;
\end{equation}
cannot separate into two factors:\cite{Mattuck:76}
\begin{equation} 
\exp\lt[t(\ke{} + \hv_s)\rt] 
\ne \exp\lt[t(\ke{})\rt] \times \exp\lt[t\hv_s\rt],\;\; \mbox{if $\lt[\ke{},\hv_s\rt]\ne 0$}.  
\end{equation}

In order to make our formulas less cluttered, especially when considering high
order diagrams, we introduce a notation where the brackets are understood:
\begin{eqnarray}
\cE_{2u}[\chi_o,\chi_u,v,v_s] = 
\hs{65}\\\nn \hs{5} 
2\smd{wr} 
\int \int d\mr_1 d\mr_2 
\int_0^{\infty} dt_1   \, 
e^{t_1\ctF_{v_s}^{1*}} \chi_r^*(\mr_1)
\hvcoroo(\mr_1)
e^{-t_1\ctF_{v_s}^1}\chi_w(\mr_1)
\chi_w^*(\mr_2) \hvcoroo(\mr_2)\chi_r(\mr_2),
\end{eqnarray}
where $\ctF_{v_s}$ is identical to $\chF_{v_s}$, except that it acts exclusively
on $\chi$; it does not act on the correlation operator $\hvcoroo$ (or its kernel
$\vcoroo$) or the complex conjugate function $\chi^*$. Similarly,
$\ctF_{v_s}^{*}$ is identical to $\chF_{v_s}$, except that it acts exclusively on
$\chi^*$; it does not act on $\vcoroo$ or $\chi$. The following identities
clarify these definitions:
\begin{eqnarray} 
\zlabel{5281}
e^{-t_1\ctF_{v_s}^1}\chi^*(\mr_1)\hvcoroo(\mr_1)\chi(\mr_1)&=&
 \chi^*(\mr_1)\hvcoroo(\mr_1)e^{-t_1\chF_{v_s}^{1}}\chi(\mr_1), \\
\zlabel{5291}
e^{-t_1\ctF_{v_s}^{1*}}\chi(\mr_1)\hvcoroo(\mr_1)\chi^*(\mr_1)&=&
\chi(\mr_1)\hvcoroo(\mr_1)e^{-t_1\chF_{v_s}^{1}}\chi^*(\mr_1).
\end{eqnarray}

In order to further compress our notation, we define the following transformed
correlation operator:
\begin{equation}
\tvcoroozvs{i} 
=e^{t_i\ctF_{v_s}^{i*}} \hvcoroo(\mr_i) e^{-t_i\ctF_{v_s}^i},
\end{equation}
where $\mz_i$ is a combined index of $\mr_i$ and $t_i$, i.e., $\mz_i =
\mr_i,t_i$. This notation yields
\begin{equation} \zlabel{3827}
\cE_{2u}[\roo,\kappa_{\roo},v,v_s] = 
\fc{1}{2} \int \int d\mr_1 d\mr_2 
\int_0^{\infty} dt_1   \, 
\tvcoroozvs{1}
\roo(\mr_1,\mr_2) 
\hvcoroo(\mr_2)
\kappa_{\roo}(\mr_2,\mr_1),
\end{equation}
where $\roo$ is given by Eq.~(\ref{7298}), while its orthogonal complement,
$\kappa_{\roo}$, is given by
\begin{equation} \zlabel{2827}
\kappa_{\roo}(\mr_1,\mr_2)= 
2 \sum_{r\in\{\chi_u\}}  \chi_{r}(\mr_1) \chi_{r}^*(\mr_2),
\end{equation}
and the sum is over the unoccupied orbitals $\{\chi_u\}$. For a complete set of
basis functions, we have
\begin{equation} \zlabel{8074}
\delta (\mr_1-\mr_2)=  \sum_{i}  \chi_{i}(\mr_1) \chi_{i}^*(\mr_2),
\;\; i \in \{\chi_o\} \cup \{\chi_u\},
\end{equation}
and therefore
\begin{equation} \zlabel{8094}
\delta (\mr_1-\mr_2) = \fc12\lt(\roo(\mr_1,\mr_2) + \kappa_{\roo}(\mr_1,\mr_2)\rt).
\end{equation}
So we can remove $\kappa_{\roo}$ from Eq.~(\ref{3827}), giving
\begin{widetext}
\begin{eqnarray} \zlabel{7629}
\cE_{2u}[\roo,v,v_s] = 
\int \int d\mr_1 d\mr_2 
\int_0^{\infty} dt_1   \, 
\tvcoroozvs{1}
\roo(\mr_1,\mr_2) 
\hvcoroo(\mr_2)
\dt(\mr_2-\mr_1)
\hs{20} \\ \hs{30} \nn \mbox{} 
-
\fc{1}{2} \int \int d\mr_1 d\mr_2 
\int_0^{\infty} dt_1   \, 
\tvcoroozvs{1}
\roo(\mr_1,\mr_2) 
\hvcoroo(\mr_2)
\roo(\mr_2,\mr_1).
\end{eqnarray}
\end{widetext}
Hence, we have $\cE_{2u}$ as an explicit functional $v$, $v_s$, and $\roo$.

In order to evaluate Eq.~(\ref{7629}), one must handle the operator $\exp(t\chF)$
acting on the Dirac delta function $\delta$, since the first term on the right
side is given by
\begin{eqnarray} \zlabel{7241}
\int \int d\mr_1 d\mr_2 \int_0^{\infty} dt_1 \, \tvcoroozvs{1}
\roo(\mr_1,\mr_2) \hvcoroo(\mr_2) \dt(\mr_2-\mr_1) 
\hs{30} \\ \nn \hs{10} 
= \int \int d\mr_1 d\mr_2 \int_0^{\infty} dt_1 \, \hvcoroo(\mr_1) \lt(e^{-t_1\chF_{v_s}^{1}}
\roo(\mr_1,\mr_2)\rt) \hvcoroo(\mr_2) \lt(e^{t_1\chF_{v_s}^{1}} \dt(\mr_2-\mr_1)\rt).
\end{eqnarray}
Using the expansion for $e^{t\chF}$ given by Eq.~(\ref{7293}), this factor will yield terms
containing the Laplacian operator $\nabla^2$ acting on $\delta$, i.e.,
$\nabla^2\delta$, as well as other higher order terms. e.g, $\nabla^4\delta$.  One
way to handle these terms is by using a basis set expansion for $\delta$, as in
Eq.~(\ref{8074}), where any complete basis can be used that satisfy the
boundary conditions. Another possibility is to use one of the integral
representations of the delta function, for example,
\begin{equation} 
\delta(\mr_1-\mr_2) = \frac{1}{8\pi^3} \int e^{i\mk\cdot(\mr_1-\mr_2)} d\mk .  
\end{equation}

Now consider the second term of Eq.~(\ref{8581}), given by the second diagram from
Fig.~\ref{9180}; evaluating this diagram yields the following expression:
\begin{equation}
\cE_{2d}[\chi_o,\chi_u,\eps] = 
2
\smd{w\in\{\chi_o\}} \smd{x\in\{\chi_o\}}
\smd{r\in\{\chi_u\}} \smd{s\in\{\chi_u\}}
\fc{1}{\eps_w-\eps_r+\eps_x-\eps_s} (rw|sx)(wr|xs),
\end{equation}
and this term can written as
\begin{eqnarray}
\cE_{2d}[\chi_o,\chi_u,\eps]  = 
2\smd{wx}\smd{rs}
(\eps_w-\eps_r+\eps_x-\eps_s)^{-1}
\hs{40}
\\ \nn
\times
\int \int \int \int d\mr_1 d\mr_2 d\mr_3 d\mr_4 \, 
\rc{12}
\chi_r^*(\mr_1) 
\chi_w(\mr_1)
\chi_s^*(\mr_2)
\chi_x(\mr_2)
\chi_w^*(\mr_3)
\chi_r(\mr_3)
\rc{34}
\chi_x^*(\mr_4)
\chi_s(\mr_4).
\end{eqnarray}
Using a generalization of Eq.~(\ref{6281}), we have
\begin{eqnarray}
\cE_{2d}[\chi_o,\chi_u,\eps]  = 
2\smd{wx}\smd{rs}
\int \int \int \int d\mr_1 d\mr_2 d\mr_3 d\mr_4  \int_0^\infty d t_{12}
\hs{30}
\\ \nn
\times
e^{t_{12}\eps_r}
e^{t_{12}\eps_s}
\rc{12}
e^{-t_{12}\eps_w}
e^{-t_{12}\eps_x}
\chi_r^*(\mr_1) 
\chi_w(\mr_1)
\chi_s^*(\mr_2)
\chi_x(\mr_2)
\chi_w^*(\mr_3)
\chi_r(\mr_3)
\rc{34}
\chi_x^*(\mr_4)
\chi_s(\mr_4).
\end{eqnarray}
Using our notation, defined by Eqs.~(\ref{5281}) and (\ref{5291}), and
Eqs.~(\ref{2827}) and (\ref{7298}), we obtain
\begin{widetext}
\begin{eqnarray} \zlabel{5381}
\cE_{2d}[\roo,v_s]  = 
\fc18 
\int \int \int \int d\mr_1 d\mr_2 d\mr_3 d\mr_4  \int_0^\infty d t_{12} 
\hs{35} \\ \hs{35} \nn 
\times
(\tilde{r}_{12}^{\hv_s})^{-1}
\roo(\mr_1,\mr_3)
\kappa_{\roo}(\mr_3,\mr_1) 
\roo(\mr_2,\mr_4)
\kappa_{\roo}(\mr_4,\mr_2)
\rc{34},
\end{eqnarray}
\end{widetext}
where the transformed electron-electron interaction term
$(\tilde{r}_{12}^{\hv_s})^{-1}$ depends on the variables $t_{12}$, $\mr_1$,
$\mr_2$, and the function $v_s$; this operator has the following form:
\begin{equation} \zlabel{4271}
\fc{1}{\tilde{r}_{12}^{\hv_s}}
=
e^{t_{12}\ctF_{v_s}^{1*}}e^{t_{12}\ctF_{v_s}^{2*}} 
\rc{12} e^{-t_{12}\ctF_{v_s}^1} e^{-t_{12}\ctF_{v_s}^2}.
\end{equation}

The $2d$ subscript indicates that $\cE_{2d}$ correspond to a second order
``direct'' diagram.  The second-order exchange term $\cE_{2x}$, from Eq.~(\ref{8581}),
is represented by the third diagram of Fig.~\ref{9180}. This diagram can be
evaluated in a similar way as the direct one $\cE_{2d}$, giving
\begin{widetext}
\begin{eqnarray} \zlabel{5385}
\cE_{2x}[\roo,v_s]  = 
-\fc{1}{16}
\int \int \int \int d\mr_1 d\mr_2 d\mr_3 d\mr_4  \int_0^\infty d t_{12}
\hs{30} \\ \nn \hs{40} 
(\tilde{r}_{12}^{\hv_s})^{-1}
\roo(\mr_1,\mr_3)
\kappa_{\roo}(\mr_3,\mr_2) 
\roo(\mr_2,\mr_4)
\kappa_{\roo}(\mr_4,\mr_1)
\rc{34}. 
\end{eqnarray}
\end{widetext}
Note that the $\cE_{2d}$ and $\cE_{2x}$ do not explicitly depend on the external
potential $v$.

In Fig.~\ref{9180}, the dummy indices that are used are denoted in the
diagrams. For example, the first diagram has $\mz_1$ for the first vertex and
$\mr_2$ for the second index, where these choices are indicated by 1 and 2 in the
figure. In order to evaluate this diagram and obtain Eq.~(\ref{3827}) directly, we
start at the bottom vertex and go backwards: The bottom vertex gives a factor of
$-\tvcoroozvs{1}$; the hole line from vertex 1 to 2 gives a factor of
$\fc{1}{2}\roo(\mr_1,\mr_2)$; the top vertex gives $-\hvcoroo(\mr_2)$; the
particle line from vertex 2 to 1 gives
$\fc{1}{2}\kappa_{\roo}(\mr_2,\mr_1)$. Furthermore, there is an additional factor
of two for the loop, since the spin state is conserved on an orientated
path.\cite{Paldus:75} Similarly, the second two diagrams from the figure can be
evaluated to give Eqs.~(\ref{5381}) and (\ref{5385}), where in both diagrams, the
lower interactions, with dummy indices $\mr_1$, $\mr_2$ and $t_{12}$, yield the
operator $(\tilde{r}_{12}^{\hv_s})^{-1}$. For the second diagam in the figure, the
direct one $\cE_{2d}$, we have the following factors: a $\fc12$, since the diagram
is symmetric, a factor of $2\times 2$ for the two loops, and a factor of
$\fc12\times\fc12\times\fc12\times\fc12$ for the two $\roo$ terms and two $\kappa$
terms. Putting this together gives $\fc12\times2\times
2\times\fc12\times\fc12\times\fc12\times\fc12=\fc18$, in agreement with
Eq.~(\ref{5381}).

Now consider the fourth diagram in Fig.~\ref{9180}, where for convenience we
denote this third order diagram by $\cE_{3d}^{\text{hh}}$, since this is a direct
diagram where the middle interaction lies on two hole lines. Evaluating this
diagram gives
\begin{equation} \zlabel{5271}
\cE_{3d}^{\text{hh}} =
2\sum_{wx}\sum_{rs}\sum_{yz}
\fc{(yr|zs)(wy|xz)(rw|sx)}
{(\eps_y-\eps_r+\eps_z-\eps_s)(\eps_w-\eps_r+\eps_x-\eps_s)},
\end{equation}
where we have two energy denominators: One corresponding to the lowest
interaction, given by $(\eps_w+\eps_x-\eps_r-\eps_s)$, and one for the middle
interaction, given by $(\eps_y+\eps_z-\eps_r-\eps_s)$, where we denote the
occupied orbitals by $w$, $x$, $y$, and $z$ and the unoccupied orbitals by $r$ and
$s$.  Using the following identity:\cite{Harris:92}
\begin{equation} \zlabel{4281}
\fc{1}{\vep_j\vep_i} = 
\int_{0}^\infty dt_2\, e^{-t_2(\vep_j-\vep_i)}
\int_{t_2}^\infty dt_1\, e^{-t_1\vep_i},
\end{equation}
the product of the two energy denominator can be written as
\begin{eqnarray} \label{9763}
\fc{1}{(\eps_y-\eps_r+\eps_z-\eps_s)(\eps_w-\eps_r+\eps_x-\eps_s)}
\hs{50} \\ \nn \hs{35}
= 
\int_{0}^\infty dt_{34}\, e^{-t_{34}(\eps_y-\eps_w+\eps_z-\eps_x)}
\int_{t_{34}}^\infty dt_{12}\, e^{-t_{12}(\eps_w-\eps_r+\eps_x-\eps_s)}.
\end{eqnarray}
Comparing the right side of this expression with the the fourth diagram in the
figure, or with Eq.~(\ref{5271}), we see that the two lowest interactions, with
the following two matrix elements: $(rw|sx)$ and $(wy|xz)$, are directly related
to the energy exponential factors: $t_{12}(\eps_r-\eps_w+\eps_s-\eps_x)$ and
$t_{34}(\eps_w-\eps_y+\eps_x-\eps_z)$ . Therefore, multiplying Eq.~(\ref{9763}) by these
two matrix elements gives
\begin{eqnarray} \label{0423}
\fc{(rw|sx)(wy|xz)}{(\eps_y-\eps_r+\eps_z-\eps_s)(\eps_w-\eps_r+\eps_x-\eps_s)}
=
\int \int \int \int d\mr_1 d\mr_2 d\mr_3 d\mr_4 
\hs{25}
\\ \nn
\hs{15}
\times
\int_{0}^\infty dt_{34} \int_{t_{34}}^\infty dt_{12}\, 
\lt(
e^{t_{12}\eps_r}e^{t_{12}\eps_s}
\rco 
e^{-t_{12}\eps_w}e^{-t_{12}\eps_x}
\chi_r^*(\mr_1) \chi_w(\mr_1) 
\chi_s^*(\mr_2) \chi_x(\mr_2)\rt) \,
\\ \nn 
\hs{30} 
\times
\lt(
e^{t_{34}\eps_w} e^{t_{34}\eps_x} 
r_{34}^{-1} 
e^{-t_{34}\eps_y} e^{-t_{34}\eps_z} 
\chi_w^*(\mr_3) \chi_y(\mr_3) 
\chi_x^*(\mr_4)\chi_z(\mr_4) \rt)
\\ \nn
= 
\int \int \int \int d\mr_1 d\mr_2 d\mr_3 d\mr_4 
\int_{0}^\infty dt_{34} \int_{t_{34}}^\infty dt_{12}
\hs{45} \\ \hs{20} \nn
\times
(\tilde{r}_{12}^{\hv_s})^{-1}
\chi_r^*(\mr_1) \chi_w(\mr_1) 
\chi_s^*(\mr_2) \chi_x(\mr_2) 
(\tilde{r}_{34}^{\hv_s})^{-1}
\chi_w^*(\mr_3) \chi_y(\mr_3) 
\chi_x^*(\mr_4) \chi_z(\mr_4), 
\end{eqnarray}
where we used Eq.~(\ref{4271}). Comparing Eqs.~(\ref{0423}) and (\ref{5271}), and
using Eqs.~(\ref{7298}) and (\ref{2827}),  gives
\begin{widetext}
\begin{eqnarray} 
\cE_{3d}^{\text{hh}}[\roo,v_s]=
\fc{1}{32}
\int \int \int \int \int \int d\mr_{1} d\mr_{2} d\mr_{3} d\mr_{4} d\mr_5 d\mr_6 
\int_{0}^\infty dt_{34} \int_{t_{34}}^\infty dt_{12}
\hs{20}\\ \hs{10} \nn
\times
(\tilde{r}_{12}^{\hv_s})^{-1}
(\tilde{r}_{34}^{\hv_s})^{-1}
r_{56}^{-1}
\roo(\mr_1,\mr_3) 
\roo(\mr_3,\mr_5) 
\kappa_{\roo}(\mr_5,\mr_1) 
\roo(\mr_2,\mr_4)
\roo(\mr_4,\mr_6)
\kappa_{\roo}(\mr_6,\mr_2).
\end{eqnarray}
\end{widetext}

Note that in this diagram the two bottom interactions, $\rco$ and $r_{34}^{-1}$,
are transformed into $(\tilde{r}_{12}^{\hv_s})^{-1}$ and
$(\tilde{r}_{34}^{\hv_s})^{-1}$, respectively, while the top interaction,
$r_{56}^{-1}$, is not transformed. Furthermore, the transformed interaction that
is next to the top, $(\tilde{r}_{34}^{\hv_s})^{-1}$, with the dummy index
$t_{34}$, gives an integration limit of $0$ to $\infty$, while the one below that,
$(\tilde{r}_{12}^{\hv_s})^{-1}$, with the dummy index $t_{12}$, has an integration
limit of $t_{34}$ to $\infty$.  By considering higher-order diagrams, and by
generalizing the identity given by Eq.~(\ref{4281}), e.g.,\cite{Harris:92}
\begin{equation}
\fc{1}{\vep_3\vep_2\vep_1} = 
\int_{0}^\infty dt_3\, e^{-t_2(\vep_3-\vep_2)}
\int_{t_3}^\infty dt_2\, e^{-t_2(\vep_2-\vep_1)}
\int_{t_2}^\infty dt_1\, e^{-t_1\vep_1},
\end{equation}
it is easily shown that the following rules apply to all diagram: 1) A factor of
$\fc12\roo$ and $\fc12\kappa_{\roo}$ is given for each hole and particle line,
respectively. 2) Interactions given by $\hvcoroo(\mr_i)$ and $r_{ij}^{-1}$ are
transformed into $\tvcoroozvs{i}$ and $(\tilde{r}_{ij}^{\hv_s})^{-1}$, unless they
are the last, or uppermost, interaction. 3) The spatial dummy indices, $\mr_1,
\mr_2, \cdots$, are integrated over. 4) A $t_i$ dummy index from $\tvcoroozvs{i}$
and a $t_{ij}$ index from $(\tilde{r}_{ij}^{\hv_s})^{-1}$ are integrated over
using the following rules to obtain their integration limits: i) If the
interaction---either $\tvcoroozvs{i}$ or $(\tilde{r}_{ij}^{\hv_s})^{-1}$---is the
second from the top---the next to the last one---the integration limit is from $0$
to $\infty$. ii) For other interactions, the integration limit is from $t_j$ to
$\infty$, if the interaction above it is $\tvcoroozvs{j}$; while, the integration
limit is from $t_{jk}$ to $\infty$, if the interaction above it is
$(\tilde{r}_{jk}^{\hv_s})^{-1}$. In addition, we also have the usual rules of
symmetry and sign factors as well as a factor of two for each loop\cite{Paldus:75}
corresponding to a sum over each spin state, $\al$ and $\be$. Also, according to
Eq.~(\ref{7682}), the one-body interaction has a negative sign associated with it,
it is $-\tvcoroozvs{1}$.

As a final example, consider the last diagram appearing in Fig.~\ref{9180}, where
we denote this arbitrary (arb) fourth order diagram by $\cE_{\text{arb}}$. Using
the rules above we get
\begin{widetext}
\begin{eqnarray} 
\cE_{\text{arb}}[\roo,v,v_s]=
-\fc{1}{32} 
\int\int\int\int\int\int\int d \mr_1 d\mr_2 d\mr_3 d\mr_4 d\mr_5 d\mr_6 d\mr_7
\int_{0}^{\infty} dt_{5} \int_{t_5}^{\infty} dt_{34} \int_{t_{34}}^{\infty} dt_{12}
\hs{2} \\ \nn 
\times
(\tilde{r}_{12}^{\hv_s})^{-1}
(\tilde{r}_{34}^{\hv_s})^{-1}
\tvcoroozvs{5}
r_{67}^{-1}
\roo(\mr_1,\mr_3)
\roo(\mr_3,\mr_6)
\kappa_{\roo}(\mr_6,\mr_1)
\roo(\mr_2,\mr_4)
\roo(\mr_4,\mr_7)
\kappa_{\roo}(\mr_7,\mr_5)
\kappa_{\roo}(\mr_5,\mr_2)\lefteqn{.}
\end{eqnarray}
\end{widetext}

Hence, using the above rules for all diagram, and Eqs.~(\ref{5279}) and
(\ref{8094})---for $\hvcoroo$ and $\kappa_{\roo}$---yields the correlation energy
$\cEco$ as an explicit functional $\roo$, $v$, and $v_s$. i.e,
$\cEco[\roo,v,v_s]$. The other two choices are $\cEco$ as an explicit functional
$\roo$, $\vcoroo$ and $v_s$, or, using Eq.~(\ref{6483}), $\cEco$ as an explicit
functional $\roo$, $v$, and $\chF_\roo$. i.e, $\cEco[\roo,v,\chF_\roo]$, where the
$v$ dependence arises from the Fock operator $\hF_\roo$, using Eq.~(\ref{9158}).

It is well known the the correlation energy cannot depend on the zeroth order
Hamiltonian $\chF_\roo$. Therefore, while the expansion $\cEco[\roo,v,\chF_\roo]$
has individual diagrams that explicitly depend of $\chF_\roo$, the entire series
does not. Certain infinite-order partial summations also yield results that are
independent of a zeroth order Hamiltonian. For example, if all diagrams are
included that correspond to a coupled cluster calculation, the final approximation
of $\cEco$ does not depend on the zeroth order Hamiltonian $\chF_\roo$, that is, we
have the following implicit functional: $\cEco[\roo,v]$.


\bibliography{ref}
\end{document}